\documentclass[
reprint,
amsmath,
amssymb,
aps,
]{revtex4-2}

\usepackage{graphicx}
\usepackage{dcolumn}
\usepackage{bm}
\usepackage{amsmath}
\usepackage{amssymb}
\usepackage{caption}
\usepackage{subcaption}
\usepackage{lineno}

\preprint{APS/123-QED}

\begin{document}

\title{Pattern formation induced by intraspecific interactions in a predator-prey system}

\author{Luciano Stucchi}
\email{stucchi\_l@up.edu.pe}
\affiliation{Universidad del Pac\'{\i}fico, Lima, Peru}
\affiliation{Grupo de Sistemas Complejos, Universidad Polit\'ecnica de Madrid, 28040 Madrid, Spain}

\author{Javier Galeano}
\affiliation{Grupo de Sistemas Complejos, Universidad Polit\'ecnica de Madrid, 28040 Madrid, Spain}

\author{Desiderio A. Vasquez}
\affiliation{Departamento de Ciencias, Secci\'on F\'{\i}sica, Pontificia Universidad Cat\'olica del Per\'u, Av. Universitaria 1801,
San Miguel, Lima 32, Peru}



\date{\today}

\begin{abstract}
Differential diffusion is a source of instability in population dynamics systems when species diffuse with different rates. Predator-prey systems show this instability only under certain specific conditions, usually requiring Holling-type functionals involved. Here we study the effects of intraspecific cooperation and competition on diffusion-driven instability in a predator-prey system with a different structure. We conduct the analysis on a generalized population dynamics that bounds intraspecific and interspecific interactions with Verhulst-type saturation terms instead of Holling-type functionals. We find that instability occurs due to the intraspecific saturation or intraspecific interactions, both cooperative and competitive. We present numerical simulations and show spatial patterns due to diffusion.

\end{abstract}


\keywords{Turing pattern \sep reaction-diffusion \sep population dynamics \sep predator-prey system \sep cooperation \sep competition}

\maketitle




\section{\label{sec:level1}Introduction}

Population ecology treats the increase and fluctuations of populations. Therefore, the purpose of these models is the quantification of the population size of the interact species. In this way, the very first works of Lotka-Volterra equations studied the predator-prey and competing species relations. However, in many of these studies, spatial variation is not considered, but it is necessary element to understand the complete ecological behavior \citep{Okubo2001}.

In particular, Turing instabilities on population dynamics has been studied thoroughly. Many authors have shown that only ecological interactions of opposite sign among species, like predator-prey or parasitism, may produce diffusion driven instability, but pure mutualism or antagonism, with the same sign in interaction between species, may not \citep{Murray2008,Okubo2001}. Although, a single Lotka-Volterra system can not generate diffusion-driven instability, modified models might. \cite{Segel1972} showed that quadratic interactions among populations are needed in order to generate Turing instability in a predator-prey system. They introduced a quadratic positive term for the prey, understood as cooperation and a quadratic negative term for the predators, interpreted as a density dependent death term. Notably, it was also shown that cooperation among predators, introduced as a quadratic expansion of the interaction term, might not produce the same effect. The authors concluded that diffusion-driven instability is caused, in predator-prey system, by self-reinforcement mechanisms acting on the prey, the destabilizers, and self-weaking mechanisms acting on the predators, the stabilizers. \cite{Bartumeus2001} also shown that Turing instability might be produced by interference among predators in an innovative way, by constructing a ratio-dependent functional response, using a DeAngelis modified model \citep{Turchin2003}. \cite{McGehee2005} and \cite{McGehee2008} presented another case, using a modified Bazykin model \citep{Turchin2003}, where diffusion-driven instability is also produced by an interference term between predators. In this case, the interference is again produced by a quadratic negative term reflecting predators interference. The authors introduced a prey dependent interaction term between species, instead of a ratio-dependent term. These results somehow contradicts what \cite{Alonso2002} showed about only ratio-dependent functionals being able of producing diffusion-driven instabilities. Ultimately, \cite{Sun2009} showed that using a quadratic term in a Holling-type II functional response also might generate Turing instabilities.

In this paper, we show that another mechanism for Turing instabilities is possible within a predator-prey system. We use a modified version of \cite{Garcia-Algarra2014b} model to show that using only quadratic interaction terms, adequately bounded by Verhulst-type saturations, may produce diffusion-driven instability. These instabilities appear whether intraspecific direct interactions are allowed or not. When intraspecific direct interactions are not present, the instability arises from the intraspecific saturation acting on the interspecific interaction. When intraspecific direct interactions are allowed, both cooperation and competition terms between predators and preys, promote the instability. All these conditions give rise to different scenarios that we explore in the following section.

\section{\label{sec:level2}The Model}

Diffusion-driven instability takes place in predator-prey systems only under special conditions upon the intraspecific coefficients \citep{Okubo2001}. For a generic reaction-diffusion system, in dimensionless form, such as:

\begin{eqnarray}
\label{eq:popmodel1}
\frac{\partial X_1}{\partial t} &=& \nabla^2 X_1 + f_1\left(X_1,X_2\right), \\
\label{eq:popmodel2}
\frac{\partial X_2}{\partial t} &=& \delta \nabla^2 X_2 + f_2\left(X_1,X_2\right),
\end{eqnarray}

\noindent it is required, according to \cite{Murray2008}, that at least the partial derivatives satisfy

\begin{eqnarray}
\label{eq:condition1}
f_{11} &+& f_{22}<0, \text{ and} \\
\label{eq:condition2}
f_{11}f_{22} &-& f_{12}f_{21}>0,
\end{eqnarray}

\noindent with $f_{ij}=\partial f_i/\partial X_j$. Here $t$ corresponds to time, the operator $\nabla^2$ indicates the Laplacian, the functions $X_i$ are the dimensionless populations of the species $i$ and the parameter $\delta$ describes the ratio between their diffusivities ($\delta = d_2/d_1$). Models with Holling-type II functionals can meet Eqs.~\eqref{eq:condition1}--\eqref{eq:condition2} requirements, but Verhulst-type functionals can not meet them \citep{Okubo2001}.

We use a generalized model of population dynamics, based on a modified version of Garc\'{\i}a-Algarra et al. population dynamics model \citep{Garcia-Algarra2014b,Stucchi19}, which bounds mutualistic behavior (otherwise unlimited) by saturation Verhulst-like terms. The functionals of a two species system are described with the following equations, in dimensionless form (see \ref{app1}):
\begin{eqnarray}
\label{eq:functionals1}
f_1\left(u_1,u_2\right) &=& \gamma u_1\left(1-q_1u_1\right. \nonumber \\
&&\left.+\left(p_{11}u_1+p_{12}u_2\right)\left(1-u_1\right)\right) \\
\label{eq:functionals2}
f_2\left(u_1,u_2\right) &=& \gamma u_2\left(s-q_2u_2\right. \nonumber \\
&&\left.+\left(p_{21}u_1+p_{22}u_2\right)\left(1-u_2\right)\right).
\end{eqnarray}

We set $\gamma=1$ to simplify the notation. Let us note that these equations include intraspecific saturation terms, on the environment ($-q_1r_1$) but also on the interspecific interactions ($1-u_1$). The system also allows the existence of $p_{ii}$, which represent direct intraspecific interactions, such as cooperation or competition, which are usually neglected. It is the presence of all these intraspecific terms what allows diffusion-driven instability in a Verhulst-type predator-prey system.

Calculating $f_{ij}$ for the stationary solutions $\bar u_{i}$, we obtain:
%
%
\begin{eqnarray}
\label{eq:jacobian1}
f_{11} &=& -\left(1+p_{12}\bar u_{2}+p_{11}\bar u_{1}^2\right), \\
\label{eq:jacobian2}
f_{12} &=& p_{12}\bar u_{1}\left(1-\bar u_{1}\right), \\
\label{eq:jacobian3}
f_{21} &=& p_{21}\bar u_{2}\left(1-\bar u_{2}\right), \\
\label{eq:jacobian4}
f_{22} &=& -\left(s+p_{21}\bar u_{1}+p_{22}\bar u_{2}^2\right).
\end{eqnarray}

Being in dimensionless equations, populations are restricted because of the scaling, within their carrying capacities, to $u_i\le 1$. Without losing generality, we set $u_1$ as the prey and $u_2$ as the predators from now on. Thus, $p_{12}<0$ and $p_{21}>0$ which mean that $f_{12}<0$ and $f_{21}>0$.

For having diffusion-driven instability, we have two possibilities according to \cite{Murray2008}. We might have $f_{11}>0$ and $f_{22}<0$ and we denote this first scenario as the autocatalytic prey. On the other hand, we also might have $f_{11}<0$ and $f_{22}>0$ and we denote this second scenario as the autocatalytic predators. Since the autocatalytic population must be the one which diffuses slower, we have that Eqs.~\eqref{eq:popmodel1}--\eqref{eq:popmodel2} is coherent with the first scenario, where $d_2>d_1$. For the second scenario, where $d_2<d_1$, instead of having $\delta\in\left]1,\infty\right[$ for Turing instability, we have $\delta\in\left]0,1\right[$. 


In the absence of terms $p_{ii}$, evaluating Eqs.~\eqref{eq:functionals1}--\eqref{eq:functionals2} for the stationary solutions $\bar u_{i}$ force that $\left(1+p_{12}\bar u_{2}\right)=\bar u_1\left(q_1+p_{12}\bar u_2\right)$ and also $\left(s+p_{21}\bar u_{1}\right)=\bar u_2\left(q_2+p_{21}\bar u_1\right)$. Since we already have that $p_{12}<0$, $f_{11}>0$ only occurs if $\mid p_{12}\mid > 1/\bar u_2$ and $\mid p_{12}\mid > q_1/\bar u_2$. This opens the possibility to a new mechanism for diffusion-driven instabilities motivated entirely by the intraspecific saturation on the interspecific interaction of Eqs.~\eqref{eq:functionals1}--\eqref{eq:functionals2}. This mechanism corresponds to the autocatalytic prey scenario. No autocatalytic predators scenario is possible, since $u_2\left(q_2+p_{21}\bar u_1\right)>0$ for any $q_2$ and $p_{21}$.

In the absence of intraspecific saturation on any interactions, either intraspecific or interspecific, Eqs.~\eqref{eq:jacobian1}--\eqref{eq:jacobian4} give the familiar result of both $f_{11},f_{22}\le 0$, that does not allow diffusion-driven instability. The case without any intraspecific saturation, not even with the environment, $f_{11},f_{22}=0$, which correspond to the classical Lotka-Volterra system \citep{Okubo2001}.

\subsection{Autocatalytic prey without intraspecific interactions}

In the absence of intraspecific interactions, i.e. for $p_{ii}=0$, we already saw that $f_{11}>0$ only if $\mid p_{12}\mid > 1/\bar u_2$ and $\mid p_{12}\mid >q_1/\bar u_2$. For this relations, we might derive,

\begin{equation}
\label{eq:autoprey_p12}
\frac{\max \left(1,q_1\right)}{\bar u_2} < \mid p_{12}\mid.
\end{equation}

On the other hand, $f_{22}<0$ always, since $\left(s+p_{21}\bar u_{1}\right)>0$ for any $p_{21}$, so no further conditions are needed.

\subsection{Autocatalytic prey with intraspecific interactions}

In this scenario, and by allowing the presence of $p_{ii}$, we might have $f_{11}>0$ only when $\left(1+p_{12}\bar u_2+p_{11}\bar u_{1}^2\right)<0$. From Eqs.~\eqref{eq:functionals1}--\eqref{eq:functionals2} evaluated for $\bar u_i$, we have,

\begin{equation}
\label{eq:autoprey_p11p12}
p_{11}\bar u_1+p_{12}\bar u_2 = \frac{q_1\bar u_1-1}{1-\bar u_1},
\end{equation}

so, for $f_{11}<0$ we need that,

\begin{equation}
\label{eq:autoprey_p11}
\frac{q_1-1}{\left(1-\bar u_1\right)^2} + p_{11} < 0.
\end{equation}

This condition allows two possible behaviors for $p_{11}$, i.e. for the intraspecific interactions of the prey. For $q_1>1$, prey must be competitive and $\mid p_{11} \mid > (q_1-1)/(1-\bar u_1)^2$. For $q_1<1$, prey might be competitive, without any restriction, or cooperative, as long as $\mid p_{11} \mid < (q_1-1)/(1-\bar u_1)^2$. The value of $q_1$ comes from their interpretation in Eqs.~\eqref{eq:popmodel1}--\eqref{eq:popmodel2}, through the transformations showed in \ref{app1}. $q_1=\left(1/c_1\right)/\left(r_1/a_1\right)$, which might be understood as the ratio between the population limit due exclusively to the resources obtained from the interspecific and intraspecific interactions, $1/c_1$, and the population limit due exclusively to the resources from the environment, $r_1/a_1$.

On the other hand, $f_{22}<0$ requires $\left(s+p_{21}\bar u_1+p_{22} \bar u_{2}^2\right)>0$. For the stationary solution, $f_2=0$ in Eqs.~\eqref{eq:functionals1}--\eqref{eq:functionals2}, we derive,

\begin{equation}
\label{eq:autoprey_p22p21}
p_{21}\bar u_1+p_{22}\bar u_2 = \frac{q_2\bar u_2-s}{1-\bar u_2},
\end{equation}

so, for $f_{22}>0$ we need that,

\begin{equation}
\label{eq:autoprey_p22}
\frac{q_2-s}{\left(1-\bar u_2\right)^2} + p_{22} > 0,
\end{equation}

which allows both cooperative and competitive predators, regardless of the sign of $s$. As long as $0<q_2-s$, predators might be cooperative, without any restriction, or they might be competitive, as long as $\mid p_{22} \mid < (q_2-s)/(1-\bar u_2)^2$. But, if $q_2-s<0$, predators must be cooperative and $\mid p_{22} \mid > (q_2-s)/(1-\bar u_2)^2$. We will see later that the intensity of this self-interaction $p_{22}$ will condition the value of the critical diffusion. 

\subsection{Autocatalytic predators}

In this other scenario, the only change is that now $f_{11}<0$ and $f_{22}>0$ are required. For the first condition, it is needed that $\left(1+p_{12}\bar u_{2}+p_{11}\bar u_1^2\right)>0$. Using what we derived on the previous section, since Eq.~\eqref{eq:autoprey_p11p12} is fulfilled again, instead of Eq.~\eqref{eq:autoprey_p11}, we have,

\begin{equation}
\label{eq:autopredators_p11}
\frac{q_1-1}{\left(1-\bar u_1\right)^2} + p_{11} > 0.
\end{equation}

Now, this condition allows both cooperative and competitive prey in the opposite direction of what happened in the previous scenario. For $q_1>1$, prey might be competitive, as long as $\mid p_{11} \mid < (q_1-1)/(1-\bar u_1)^2$ and cooperative without any restriction. On the other hand, if $q_1<1$, prey must be cooperative and $p_{11} > \mid q_1-1 \mid (1-\bar u_1)^2$. We will also see that the intensity of $p_{11}$ will determine the value of the critical diffusion.

On the other hand, for $f_{22}>0$, we also use Eq.~\eqref{eq:autoprey_p22p21} and now, instead of Eq.~\eqref{eq:autoprey_p22}, we have,

\begin{equation}
\label{eq:autopredators_p22}
\frac{q_2-s}{\left(1-\bar u_2\right)^2} + p_{22} < 0.
\end{equation}

For $q_2-s>0$, $p_{22}<0$ and predators must be competitive, but additionally, $\mid p_{22} \mid > (q_2-s)/(1-\bar u_2)^2$. But, when $q_2-s<0$, predator might be competitive, without any restriction, or cooperative, as long as $p_{22} < \mid q_2-s \mid /(1-\bar u_2)^2$

All the conditions derived in the last two scenarios are only established to see the possible ecological regimes, i.e. the signs allowed on $p_{ii}$, that are valid in order to produce diffusion-driven instability. Since $\bar u_i$ are functions of $p_{ii}$, no simple relation can be obtained from Eqs.~\eqref{eq:autoprey_p11p12}--\eqref{eq:autopredators_p22}. The same applies to the first scenario, about $p_{12}$ and its relation with $q_1$ and $\bar u_2$. This can be seen in the Results.

\subsection{Diffusion-driven instability}

Diffusion-driven instabilities require, according to \cite{Murray2008}, that conditions \eqref{eq:condition1}--\eqref{eq:condition2} change into,

\begin{eqnarray}
\label{eq:newconditions1}
\delta f_{11} &+& f_{22}>0, \\
\label{eq:newconditions2}
\left(\delta f_{11}+f_{22}\right)^2 &-& 4\delta \left(f_{11}f_{22}-f_{12}f_{21}\right)>0.
\end{eqnarray}
\\
These conditions explain the reason why, considering the autocatalytic prey scenario with intraspecific interactions, in Eqs.~\eqref{eq:autoprey_p22} $p_{22}$ was an indicator of the critical diffusion $\delta_c$. This applies also for the autocatalytic predators scenario, except that in that case, it is $p_{11}$ the parameter that acts as an indicator of the $\delta_c$. From \cite{Murray2008} it is required that $\delta_c$ follows, 

\begin{equation}
\label{eq:deltac}
\delta_c^2f_{11}^2+2\delta_c\left(2f_{12}f_{21}-f_{11}f_{22}\right)+f_{22}^2=0.
\end{equation}


This means that, at least, $\mid f_{22}/f_{11}\mid <\delta_c$ for the autocatalytic prey scenario. In other words, given a $f_{11}$, the greater the cooperation of the predators, the greater critical diffusion will be needed to get a diffusion-driven instability. In the autocatalytic predators scenario we have that $\mid f_{22}/f_{11}\mid >\delta_c$, i.e. the inverse dependence is needed between them in order to get the critical diffusion.

We test for diffusion-driven instability using the non-dimensionless system. We introduce small perturbations to the homogeneous stationary solutions of the system, given by $\bar X_i$. Perturbations are introduced, as functions of fixed wavelength of the form $X_i=\bar X_i+X'_ie^{\lambda t}e^{ikz}$, into Eqs.~\eqref{eq:popmodel1}--\eqref{eq:popmodel2} and neglecting the non-linear terms \citep{Stucchi13}. This gives a set of two equations relating the eigenvalues $\lambda$ with the wavenumber $k$. This constitutes a dispersion relation from which the stability of the system can be verified. We present this relation on the following section, along with the numerical solutions of the nonlinear system.

\section{\label{sec:level3}Results}

\subsection{Linear stability analysis}

We tested the stability of the homogeneous stationary solution by replacing some test values for $c_1$ and $b_{12}$ in the autocatalytic prey scenario without intraspecific interactions, and some test values for $s$ and $p_{ij}$ in both scenarios with intraspecific interactions. Eigenvalues $\lambda$ were obtained as function of wavenumber $k$. $Re\left(\lambda\right)$ change from negative to positive for a certain values of $k$, indicating the cases where a small perturbation with wavelength $2\pi /k$ will not vanish. Instead, those perturbations will grow and will make the system unstable on a linear approximation; the system will stabilize itself by the nonlinear terms.

\begin{figure*}[htbp]
\centering\includegraphics[width=1.0\linewidth]{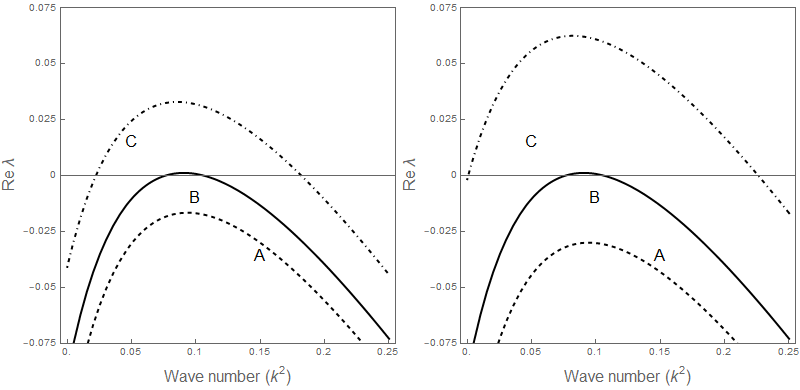}
\caption{Autocatalytic prey scenario without intraspecific interactions. Effects of predation intensity (left) and intraspecific saturation (right), which are two aspects of the interspecific relation between prey and predators, on the dependence of the real part of the eigenvalue $\lambda$ on the wavenumber $k$. We plotted the deviations from the values corresponding to Table~\ref{table:autopreywo}, which are the curves B. On the left, we set $b_{12}=-0.001015$ for A and $b_{12}=-0.001005$ for C. On the right, we set $c_1=0.00195$ for A and $c_1=0.00205$ for C. In both cases, $\delta=182$. Lower absolute values of predation intensity and higher intraspecific saturation benefit the instability of the system.}
\label{fig:autopreywolk}
\end{figure*}

For the scenario of the autocatalytic prey without intraspecific interactions, we tested the case where both species have a positive dependence on the environment, i.e. $s>0$. In Figure~\ref{fig:autopreywolk}, we show the effects of parameters $c_1$ and $b_{12}$ in the instability of the system. We see, on the left, that instability is promoted with lower values of $\mid b_{12}\mid$, i.e. with less effects on the prey by the predators and, on the right, with greater values of $c_1$, i.e. with a higher intraspecific saturation. However, this effect reaches a point where the system may become intrinsically unstable (see curve C on the right) and no diffusion-driven instability might be generated.

\begin{figure*}[htbp]
\centering\includegraphics[width=1.0\linewidth]{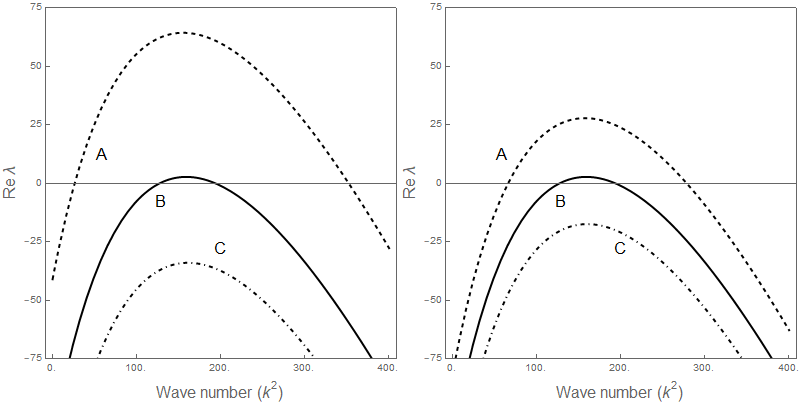}
\caption{Autocatalytic prey scenario with intraspecific interactions. Effects of cooperation of prey (left) and competition of predators (right) on the dependence of the real part of the eigenvalue $\lambda$ on the wavenumber $k$. We plotted the deviations from the values corresponding to Table~\ref{table:autoprey}, which are the curves B. On the left, we set $b_{11}=0.0018996$ for A and $b_{11}=0.0019004$ for C. On the right, we set $b_{22}=-0.0018996$ for A and $b_{22}=-0.0019004$ for C. In both cases, $\delta=19$. Competition in predators and cooperation in prey promotes the stability in both cases.}
\label{fig:autopreylk}
\end{figure*}

\begin{figure*}[htbp]
\centering\includegraphics[width=1.0\linewidth]{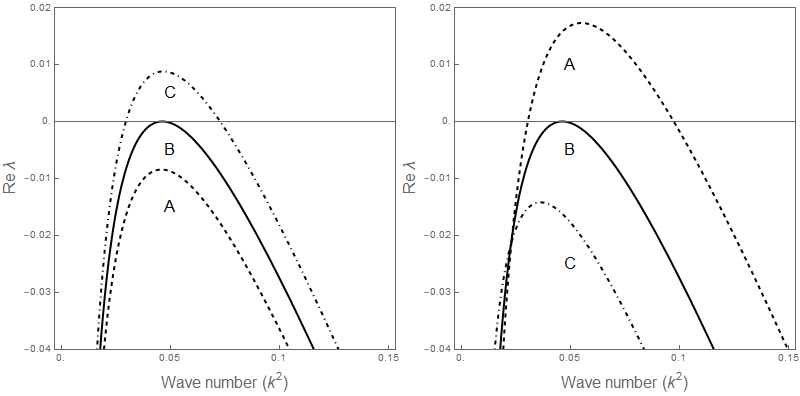}
\caption{Autocatalytic prey scenario with intraspecific interactions. Effects of cooperation of prey (left) and cooperation of predators (right) on the dependence of the real part of the eigenvalue $\lambda$ on the wavenumber $k$. We plotted the deviations from the values corresponding to Table~\ref{table:autoprey2}, which are the curves B. On the left, we set $b_{11}=0.0015$ for A and $b_{11}=0.0016$ for C. On the right, we set $b_{22}=0.000965$ for A and $b_{22}=0.001035$ for C. In both cases, $\delta=45$. Cooperation in predators promotes the stability of the system, while cooperation in prey promotes its instability.}
\label{fig:autoprey2lk}
\end{figure*}

To explore the scenario of autocatalytic prey with intraspecific interactions, we tested the case where both species have a positive dependence on the environment, i.e. $s>0$, and where predators compete and cooperate among themselves. In Figure~\ref{fig:autopreylk}, we show the effects of both competition of predators and cooperation of prey in the instability of the system. We see, on the left, that lower competition among predators promotes a greater instability in the system. But, on the right, we also see that lower cooperation of prey promotes also a greater instability. In Figure~\ref{fig:autoprey2lk}, we show the effects of both cooperation of predators and cooperation of prey in the instability of the system. On the left, we see the influence of cooperation in prey and how it promotes the system instability, while on the right, we see the influence of cooperation in predators, and how it promotes the stability instead.

\begin{figure*}[htbp]
\centering\includegraphics[width=1.0\linewidth]{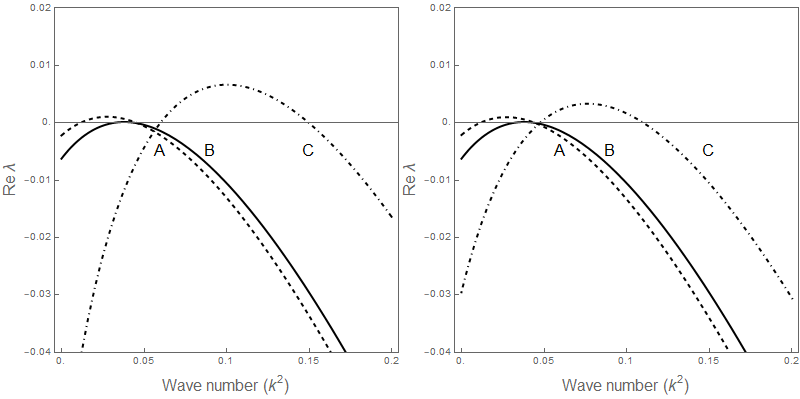}
\caption{Autocatalytic predators scenario. Effects of cooperation of prey (left) and cooperation of predators (right) on the dependence of the real part of the eigenvalue $\lambda$ on the wavenumber $k$. We plotted the deviations from the values corresponding to Table~\ref{table:autopredators}, which are the curves B. On the left, we set $b_{11}=0.001909985$ for A and $b_{11}=0.0019105$ for C. On the right, we set $b_{22}=0.00549915$ for A and $b_{22}=0.00551$ for C. In both cases, $\delta=0.357$. Cooperation in predators and prey promotes the instability in both cases, although only greater cooperation may guarantee diffusion-driven instability, since lower cooperation values (curves A) makes the system intrisincally unstable.}
\label{fig:autopredatorslk}
\end{figure*}

For the autocatalytic predators scenario, we tested the case where $s<0$, which means that prey have a positive dependence on the environment, but the predators do not. Also, besides cooperative predators, we use cooperative prey. In Figure~\ref{fig:autopredatorslk}, we show the effects of both cooperations in the instability of the system. Higher cooperation in both populations promotes Turing pattern formation, but lower cooperation also allows the system to become unstable. Continuing to lower the cooperation further results in a steady state that is unstable even without diffusion.

\begin{figure}[htbp]
\centering
\includegraphics[width=0.5\textwidth]{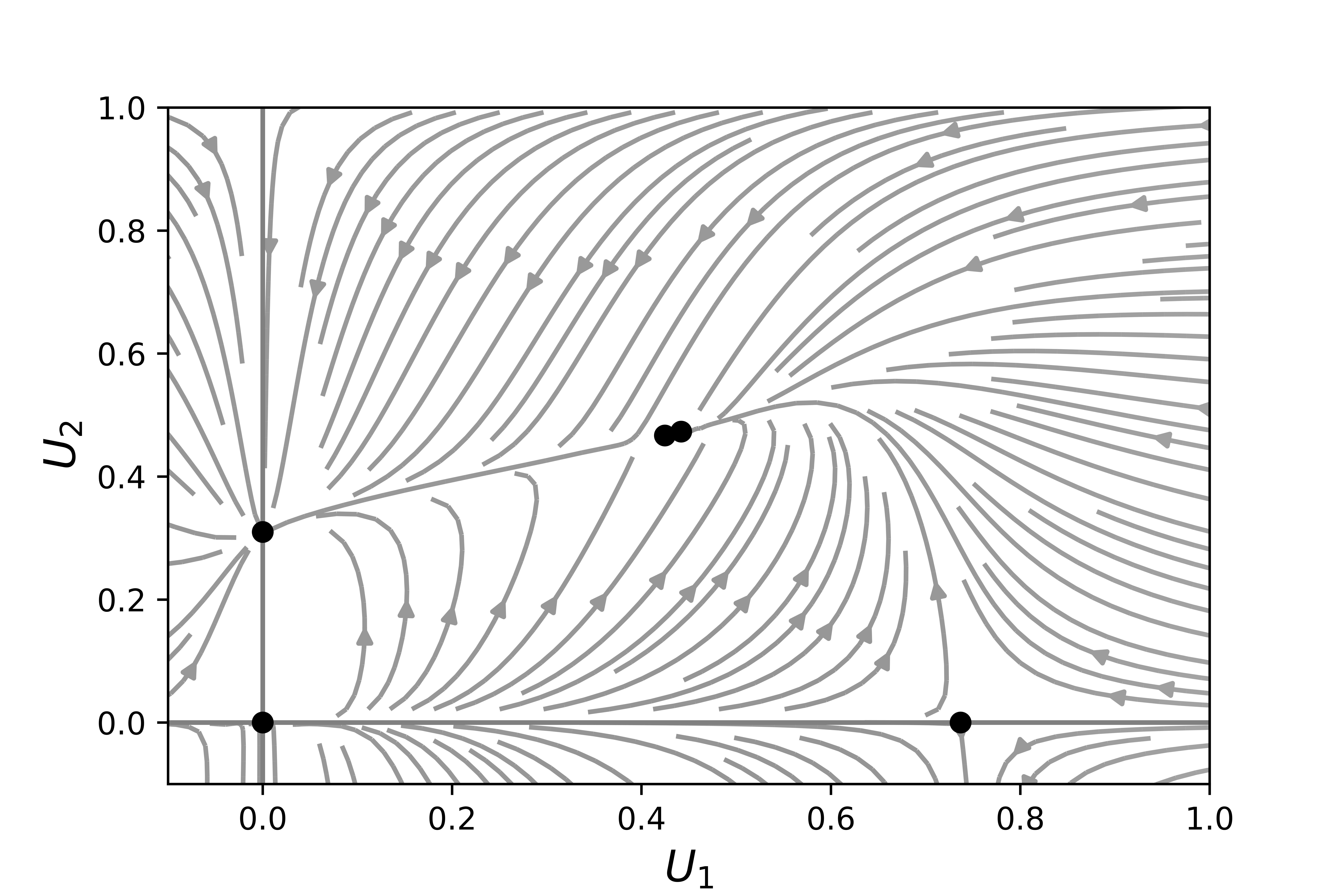}
\caption{Phase space of $X_i$ for the autocatalytic prey with intraspecific interactions. Values of the parameters correspond to those of Table~\ref{table:autoprey}. The stable solution that allows diffusion-driven instability has another unstable solution right next to it, a pair corresponding to a saddle-node bifurcation.}
\label{fig:autostreams}
\end{figure}

All solutions we tested were pairs corresponding to saddle-node bifurcations, because they converge and disappear when parameters change \citep{Strogatz1994}. It is interesting to see also that, when parameters change the other way around and solutions diverge, the diffusion-driven instability is lost. Figure \ref{fig:autostreams} shows the phase space of the autocatalytic prey scenario and values correspond to those of Table~\ref{table:autoprey}.


The scenarios discussed here reflect some differences with other previous attempts to find diffusion-driven instability in predator-prey systems with cooperative prey. \cite{Levin1976} discussed a predator-prey model without saturations and with only the prey depending on the environment. For them, prey were cooperative and predators competitive. Their model required, for diffusion-driven instability to occur, that,
\begin{eqnarray}
\label{eq:levinconditions}
p_{21} &>& p_{11}, \nonumber \\
\mid p_{21}p_{12}\mid &>&\mid p_{11}p_{22}\mid , \nonumber \\
\delta_c=\biggl(\sqrt{\frac{p_{12}}{p_{22}}}&-&\sqrt{\frac{p_{12}}{p_{22}}-\frac{p_{11}}{p_{21}}}\biggr)^{-2},
\end{eqnarray}
\noindent in the specific scenario of autocatalytic prey with intraspecific interactions. This conditions are not met neither with values in Table~\ref{table:autoprey} or Table~\ref{table:autoprey2} .

\subsection{Numerical simulations}

We solve the nonlinear system by carrying out a numerical simulation of Eqs.~\eqref{eq:popmodel1}--\eqref{eq:popmodel2}. Since only two possible patterns may arise in a one dimensional system, which are identical or inverse \cite{Murray2008}, we use values of Table~\ref{table:autoprey2} and Table~\ref{table:autopredators} to test both scenarios with intraspecific interactions, the autocatalytic prey and the autocatalytic predators. We chose periodic boundary conditions along a one-dimensional space with cell width of size $\Delta z=0.1$ spatial units, where both species $X_i$ evolve. We use a simple Euler method with a time step of $\Delta t=0.0001$, which we tested to be accurate. Initial conditions where set with small random perturbations around the homogeneous stationary solutions $\bar X_i$. Computations where carried out for enough time in order to reach a steady pattern.

\begin{figure*}[htbp]
\centering\includegraphics[width=1.0\linewidth]{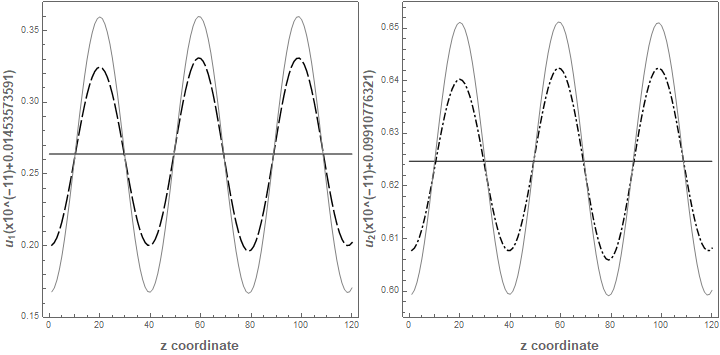}
\caption{Numerical simulations of the nonlinear system. The curves represent the dimensionless population of the autocatalytic prey scenario with intraspecific interactions and with parameter values of Table~\ref{table:autoprey2}. The straight line is drawn on the homogeneous stationary solutions. The dashed and dot-dashed lines represent $u_i$ when they reach their constant values, while the solid gray lines represent an intermediate previous state (t=120 and t=100 in Figure~\ref{fig:autopreytime}). Although amplitudes differ significantly, both populations follow the same dynamics, as expected for the autocatalytic prey scenario. We used $\delta=45$.}
\label{fig:autopreysimu}
\end{figure*}
\begin{figure*}[htbp]
\centering\includegraphics[width=1.0\linewidth]{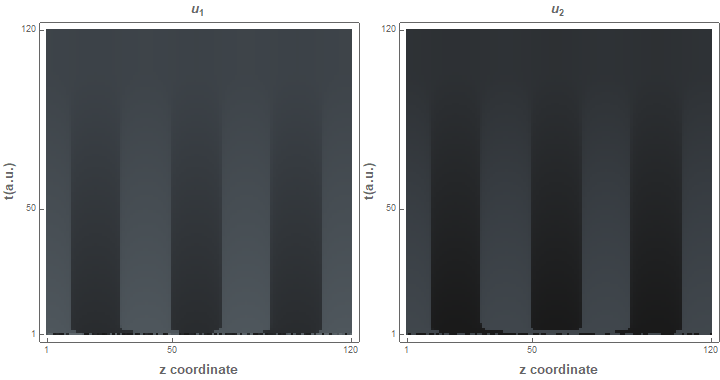}
\caption{Spatial patterns of the nonlinear system over time. The shadows represent higher (darker) or lower (lighter) values of $u_i$. The vertical axis represent the time in a.u. while the horizontal axis represent the space. Patterns corresponding to t=120 and t=100 are plotted in Figure~\ref{fig:autopreysimu}. The pattern is reached quickly, but slowly fade away until it reaches fixed values. We used parameter values of Table~\ref{table:autoprey2} and $\delta=45$.}
\label{fig:autopreytime}
\end{figure*}

For the autocatalytic prey scenario with parameter values of Table~\ref{table:autoprey2}, diffusion-driven instability appears with $\delta=45$. The corresponding wavelength of the fastest growth for this diffusion is $29.24$ spatial units. As we use a grid of cells with $\Delta z=0.1$ spatial units, it is expected to have a pattern of three or four peaks in a length of $120$ spatial units. Although an almost uniform pattern of three peaks form for both populations, their amplitudes reduce constantly until reaching a fixed value of $1.35\times 10^{-12}$ for $u_1$ and $3.65\times 10^{-13}$ for $u_2$. Both dimensionless populations show the same pattern, given that in an autocatalytic prey scenario, both species follow the same dynamics. This is shown in Figure~\ref{fig:autopreysimu}. We also show the time evolution of the pattern in Figure~\ref{fig:autopreytime}.

\begin{figure*}[htbp]
\centering\includegraphics[width=1.0\linewidth]{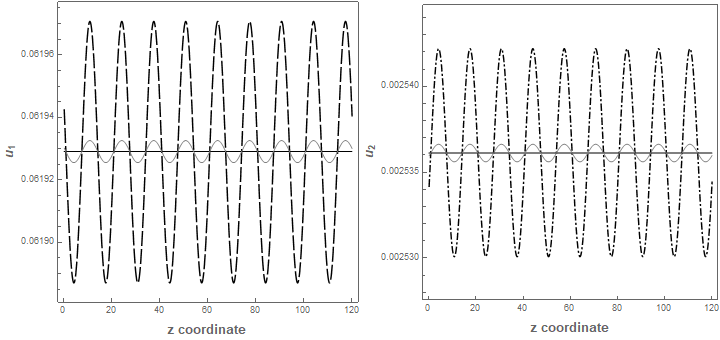}
\caption{Numerical simulations of the nonlinear system. The curves represent the dimensionless population of the autocatalytic predators scenario and with parameter values of Table~\ref{table:autopredators} and $b_{11}=0.001915$. We use $\delta=0.474$. The straight line is drawn on the homogeneous stationary solution. The dashed and dot-dashed lines represent $u_i$ when simulation was stopped, while the solid gray lines represent an intermediate previous state (t=303 and t=202 in Figure~\ref{fig:autopredatorstime}). Amplitudes are different and both populations follow the inverse dynamics, as expected for the autocatalytic prey scenario. The pattern is not stable and grows indefinitely.}
\label{fig:autopredatorssimu}
\end{figure*}
\begin{figure*}[htbp]
\centering\includegraphics[width=1.0\linewidth]{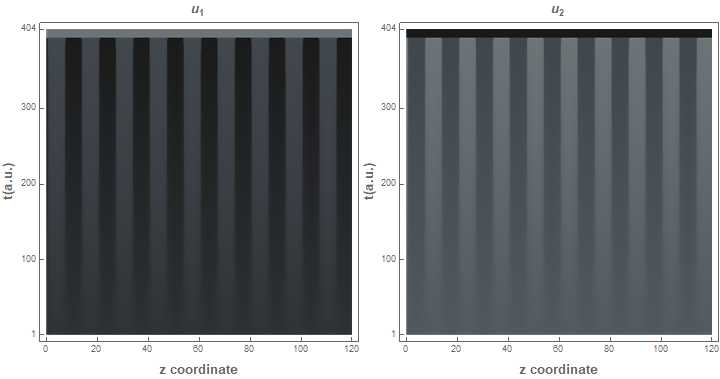}
\caption{Spatial patterns of the nonlinear system over time. The shadows represent higher (darker) or lower (lighter) values of $u_i$. The vertical axis represent the time in a.u. while the horizontal axis represent the space. Patterns corresponding to t=303 and t=202 are plotted in Figure~\ref{fig:autopredatorssimu}. The pattern is reached quickly, but it slowly increases to higher amplitudes, until it collapses in another stationary solution, a partial extinction of $u_1$. We used parameter values of Table~\ref{table:autopredators} with $b_{11}=0.001915$ and $\delta=0.474$.}
\label{fig:autopredatorstime}
\end{figure*}

For the autocatalytic predators scenario we conducted two different tests. First, we used parameter values of Table~\ref{table:autopredators} and $b_{11}=0.001915$. Diffusion-driven instability appears with $\delta=0.474$. The corresponding wavelength of the fastest growth for this diffusion is $13.49$ spatial units. With $\Delta z=0.1$ spatial units as the cell width, it is expected to have a pattern of nine peaks in a length of $120$ spatial units. Here, the dimensionless populations show an inverse pattern, given that in an autocatalytic predators scenario, species follow the opposite dynamics. This is shown in Figure~\ref{fig:autopredatorssimu}. The spatial pattern is formed with the corresponding wavelength of the Turing instability, but its amplitude continue growing indefinitely. We show the pattern at two different times in Figure~\ref{fig:autopredatorssimu} and the time evolution in Figure~\ref{fig:autopredatorstime}.

\begin{figure*}[htbp]
\centering\includegraphics[width=1.0\linewidth]{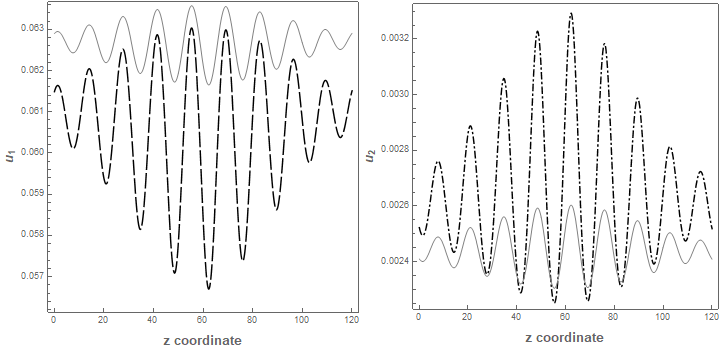}
\caption{Numerical simulations of the nonlinear system. The curves represent the dimensionless population of the autocatalytic predators scenario and with parameter values of Table~\ref{table:autopredators} and $b_{11}=0.001911$. We use $\delta=0.392$. The homogeneous stationary solution is not drawn because it is located offside of the axis ($\bar u_1=0.0677577, \bar u_2=0.00191823$). The dashed and dot-dashed lines represent $u_i$ just before it reaches higher and lower enough values to move onto another stationary solution, a partial extinction of $u_1$. The gray solid lines represent them a few step earlier (t=80 and t=75 in Figure~\ref{fig:autopredators2time}). Amplitudes are different and both populations follow the inverse dynamics, as expected for the autocatalytic prey scenario. The pattern is not stable and grows indefinitely.}
\label{fig:autopredators2simu}
\end{figure*}

\begin{figure*}[htbp]
\centering\includegraphics[width=1.0\linewidth]{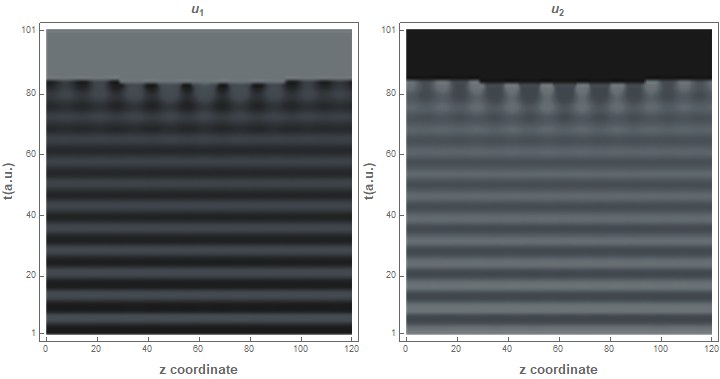}
\caption{Spatial patterns of the nonlinear system over time. The shadows represent higher (darker) or lower (lighter) values of $u_i$. The vertical axis represent the time in a.u. while the horizontal axis represent the space. Patterns corresponding to t=80 and t=75 are plotted in Figure~\ref{fig:autopredators2simu}. The pattern is reached quickly, but it slowly increases to higher amplitudes, until it collapses in another stationary solution, a partial extinction of $u_1$. We used parameter values of Table~\ref{table:autopredators} with $b_{11}=0.001911$ and $\delta=0.392$.}
\label{fig:autopredators2time}
\end{figure*}

We also test the case when $b_{11}=0.001911$ and the other parameters where those of Table~\ref{table:autopredators}. Diffusion-driven instability appears with $\delta=0.392$ and the corresponding wavelength of the fastest growth for this diffusion is $18.21$ spatial units. For a length of $120$ spatial units, seven peaks would be expected, but we obtained a pattern with nine. This pattern is unstable, and not only its amplitude grows, as in Figure~\ref{fig:autopredators2simu} but it oscillates around the stationary solution, as it can be seen in the first steps of Figure~\ref{fig:autopredators2time}. Eventually, the amplitudes reaches another basin and the populations go to another stationary solution, the partial extinction of $u_1$.


\section{\label{sec:level4}Conclusions}

Here we studied the generation of patterns from intraspecific interactions, which are usually neglected in most ecological models or which are introduced ad hoc to study specific cases. \cite{Lorenz1981} observed that, among animal species, intraspecific direct interactions act as inhibitory or autocatalytic mechanisms. When individuals behave aggressively among them, this behavior promotes their dispersion across the available territory. On the other hand, when the same individuals cooperate, gregarious behaviors appeared. When these species are involved in an ecological system, both mechanism might couple and diffusion-driven instabilities arise.

In this work, we have shown that intraspecific interactions in a predator-prey system might lead to diffusion-driven instabilities. These intraspecific interactions can be positive (cooperation) or negative (competition), they can act on the predators or on the prey, or even they can be direct (being an active interaction) or indirect (acting as a saturation). This means that they are not as limited as some previous studies pointed out \cite{Segel1972,Bartumeus2001,McGehee2005,McGehee2008,Alonso2002,Sun2009}. In the absence of intraspecific direct interactions (terms $b_{ii}X_i$), saturation acting on the prey relation with environment resources might cause instability driven by diffusion as long as Eq.~\eqref{eq:autoprey_p12} holds. This mechanism leads to an autocatalytic prey scenario. No such mechanism exists for predators. When intraspecific direct interactions are present, Turing patterns might arise either with autocatalytic prey or autocatalytic predators, with conditions Eqs.~\eqref{eq:autoprey_p11p12}--\eqref{eq:autopredators_p22} that allow them to be cooperative or competitive, regardless of the scenario.

We have shown with numerical simulations that instabilities give rise to spatial patterns that might be identical for both species, in the autocatalytic prey scenario, such as Figure~\ref{fig:autopreysimu}, or inverse, in the autocatalytic predators scenario, such as Figure~\ref{fig:autopredatorssimu} or Figure~\ref{fig:autopredators2simu}. Spatial patterns are only stable in the autocatalytic prey scenario, reaching a fixed amplitude lower than the originally reached. For the autocatalytic predators scenario, initial random perturbations grow continuously, or they show oscillatory patterns of growing amplitude around the stationary solution. Their amplitudes grow until populations reach the basin of an stable stationary solution. Although unstable Turing patterns are known, specially around Hopf bifurcations \citep{Vastano1988,Pearson1993}, here we found them around a saddle-node bifurcation. \cite{Camara2016} found that Turing instabilities around a saddle-node bifurcation led to stationary Turing patterns, the opposite of what we have found.

Since ecological models deal with living species that inhabit spatial domains, patterns arising from diffusion mechanisms are relevant to a better understanding of the behavior of populations. As \cite{Lorenz1981} pointed out, intraspecific behavior is a well known mechanism to disperse or concentrate individuals. Therefore, pattern formation due to intraspecific interactions should be considered as part of the ecological behaviors that species show among their interactions.

\section*{Acknowledgements}

This work was supported by Ministry of Economy and Competitiveness of Spain (research project MTM2015-63914-P).

\appendix

\section{Nondimensionalization of the population dynamics system}
\label{app1}

In this work, we added a diffusion term to a generalized version of the population dynamics model of \cite{Garcia-Algarra2014b}, denoted as,

\begin{eqnarray}
\label{eq:eqapp1}
\frac{\partial X_i}{\partial t} &=& d_i \nabla^2 X_i +X_i\left(r_i-a_iX_i\right. \nonumber \\
&&\left.+\left(b_{ii}X_i+b_{ij}X_j\right)\left(1-c_iX_i\right)\right).
\end{eqnarray}
\\
We used the following transformation,

\begin{align*}
z^*&=z/L & t^*&=t(d_1/L^2)  & \nabla^{*2}&=\nabla^{2}/L^2\\
\delta&=d_2/d_1 & \gamma&=r_1L^2/d_1 & s&=r_2/r_1\\
u_i&=c_iX_i & q_i&=a_i/(c_ir_1) & p_{ij}&=b_{ij}/(c_jr_1)
\end{align*}

and dropped the $^*$ in order to get Eqs.~\eqref{eq:functionals1}-\eqref{eq:functionals2}.


\section{Numerical values}
\label{app2}

Here we present the numerical values used in the simulations. We presented the values according to the population equations instead of the dimensionless system, since the latter can be derived from the transformation described in \ref{app1}.

\begin{table}[ht]
\caption{Numerical values used in the simulations shown in Figure \ref{fig:autopreywolk}, which corresponds to the autocatalytic prey scenario without intraspecific interactions.}
\centering
\begin{tabular}{r l}
\\
\hline
\textbf{Parameters} & \textbf{Numerical values}\\
\hline
$r_1$ & 0.1 \\
$r_2$ & 0.01 \\
$b_{11}$ & 0.0 \\
$b_{12}$ & -0.00101 \\
$b_{21}$ & 0.015 \\
$b_{22}$ & 0.0 \\
$a_1$ & 0.00001 \\
$a_2$ & 0.005 \\
$c_1$ & 0.002 \\
$c_2$ & 0.005 \\
\hline
\label{table:autopreywo}
\end{tabular}
\end{table}

\begin{table}[ht]
\caption{Numerical values used in the simulations shown in Figure~\ref{fig:autopreylk} and Figure~\ref{fig:autostreams}, which correspond to the autocatalytic prey scenario with intraspecific interactions.}
\centering
\begin{tabular}{r l}
\\
\hline
\textbf{Parameters} & \textbf{Numerical values}\\
\hline
$r_1$ & 0.0001\\
$r_2$ & 0.6  \\
$b_{11}$ & 0.0019 \\
$b_{12}$ & -0.00075 \\
$b_{21}$ & 0.00091 \\
$b_{22}$ & -0.0019 \\
$a_1$ & 0.0005 \\
$a_2$ & 0.000625 \\
$c_1$ & 0.001251 \\
$c_2$ & 0.001 \\
\hline
\label{table:autoprey}
\end{tabular}
\end{table}

\begin{table}[ht]
\caption{Numerical values used in the simulations shown in Figure~\ref{fig:autoprey2lk}, which corresponds to the autocatalytic prey scenario with intraspecific interactions.}
\centering
\begin{tabular}{r l}
\\
\hline
\textbf{Parameters} & \textbf{Numerical values}\\
\hline
$r_1$ & 0.9 \\
$r_2$ & 0.00001 \\
$b_{11}$ & 0.00155 \\
$b_{12}$ & -0.001 \\
$b_{21}$ & 0.00075 \\
$b_{22}$ & 0.001 \\
$a_1$ & 0.001 \\
$a_2$ & 0.001 \\
$c_1$ & 0.0001 \\
$c_2$ & 0.0001 \\
\hline
\label{table:autoprey2}
\end{tabular}
\end{table}

\begin{table}[ht]
\caption{Numerical values used in the simulations shown in Figure~\ref{fig:autopredatorslk} which corresponds to the autocatalytic predators scenario. In Figures~\ref{fig:autopredatorssimu}-\ref{fig:autopredatorstime} we change $b_{11}=0.001915$ and in Figures~\ref{fig:autopredators2simu}-\ref{fig:autopredators2time} we used $b_{11}=0.001911$.}
\centering
\begin{tabular}{r l}
\\
\hline
\textbf{Parameters} & \textbf{Numerical values}\\
\hline
$r_1$ & 0.02999 \\
$r_2$ & -0.090151 \\
$b_{11}$ & 0.00191 \\
$b_{12}$ & -0.0023515 \\
$b_{21}$ & 0.00105 \\
$b_{22}$ & 0.0055 \\
$a_1$ & 0.0021 \\
$a_2$ & 0.0005 \\
$c_1$ & 0.001 \\
$c_2$ & 0.0005 \\
\hline
\label{table:autopredators}
\end{tabular}
\end{table}

\clearpage

\section*{References}

\bibliography{Diffusion_paper.bib}

\end{document}